# Structural Disorder and Electronic Structure of Sr(Ti$_x$Fe$_{1-x}$)O$_{3-x/2}$ Solid Solutions: A Computational Framework

Bin Ouyang[1], Tim Mueller[2], Nicola H. Perry[3], N. R. Aluru[1,4,5], Elif Ertekin[1,3,4]

1. National Center for Supercomputing Applications, University of Illinois at Urbana–Champaign, 1205 W Clark St, Urbana 61801, USA
2. Department of Materials Science and Engineering, Johns Hopkins University, 3400 North Charles Street, Baltimore, Maryland 21218, USA
3. International Institute for Carbon Neutral Energy Research (WPI-I2CNER), Kyushu University, 744 Motooka, Nishi-ku, Fukuoka 819-0395, Japan
4. Department of Mechanical Science and Engineering, University of Illinois at Urbana-Champaign, 1206 W Green St, Urbana 61801, USA
5. Beckman Institute for Advanced Science and Technology, University of Illinois at Urbana-Champaign, Urbana, Illinois 61801, USA

# Abstract

Several mixed ionic/electronic conductors (MIECs) used as fuel or electrolysis cell electrodes may be thought of as solid solution mixtures between perovskite oxides and ordered oxygen vacancy compounds. For example, the model MIEC $Sr(Ti_{1-x}Fe_x)O_{3-x/2+\delta}$ (STF) can be described as a mixture of the perovskite $SrTiO_3$ and the brownmillerite $Sr_2Fe_2O_5$. The large configurational space for these non-dilute, disordered alloys historically has hindered direct atomic scale modeling, which precludes in-depth understanding and predictive analysis. In this work, we present a cluster expansion framework to describe the energetics of the disordered STF alloy within the full solid solution composition space $Sr(Ti_{1-x}Fe_x)O_{3-x/2}$ ($0<x<1$, $\delta=0$). Cluster expansion Monte Carlo (CEMC) simulations are performed to determine the lowest energy atomic configurations and investigate the origin and degree of lattice disorder. Using realistic configurations obtained from CEMC for different temperaturees, the electronic structure of the solution at different stoichiometry are examined to understand their electronic structure, band gap, and optical properties, and compared and contrasted to those of hypothetical ordered structures. The evolution of the band gap and optical absorption with composition predicted using our atomic model is consistent with experiment. Meanwhile, band edge analysis elucidates that electronic transport within the alloy benefits from the simultaneous presence of Fe/Ti disorder on the B cation sublattice and a tendency for oxygen vacancies to cluster around Fe atoms. Using $SrTiO_3/Sr_2Fe_2O_5$ alloy as an example, the modeling framework adopted here can be extended to other MIEC materials.

# 1. Introduction

Mixed ionic electronic conductors (MIECs) exhibiting large electronic and oxygen-ion conductivity are important across a variety of solid-state electrochemical devices, including solid oxide fuel and electrolysis electrodes, oxygen separation membranes, oxygen sensors, and catalysts[1-7]. The $SrTi_{1-x}Fe_xO_{3-y}$ alloy[8-9] (referred to as STF) is a classic example of a complex MIEC alloy[8-15]. The STF composition space forms a continuous solid solution between $0 < x < 1$ that, depending on the Ti/Fe composition and thermodynamic environment, exhibits large, variable ionic and electronic conductivity[8-9]. This makes the STF solid solution technologically important for several practical applications, especially if control over properties can be achieved by tuning composition, oxygen richness/deficiency, and configurations.

Understanding the configurations, electronic structure, and transport properties of STF remains a challenge, and a unified picture of the structure/property relations is still missing. This is due to the difficulty in achieving realistic atomic-scale representations of the configurations of the non-dilute, disordered solution, rendering mechanistic understanding and predictive modeling challenging. The goal of this work is to introduce a computational framework that correlates composition, alloy configurations, electronic structure, and optical properties. We present a self-consistent description of the atomic-scale configurations across the full composition space $0 < x < 1$ based on a cluster expansion model. The cluster coefficients are fitted to density functional theory calculations to describe the configurational energetics, and the resulting model is used to establish in detail the effects of composition and structure on the electronic and optical properties.

In contrast to previous studies in which dilute-solution Fe-doped SrTiO$_3$ is used as a reference framework to understand STF[16-20], we adopt a formalism where the STF alloy is considered as a continuous solid solution for $0 < x < 1$ between end members SrTiO$_3$ (perovskite) and Sr$_2$Fe$_2$O$_5$ (brownmillerite)[8], with composition SrTi$_{1-x}$Fe$_x$O$_{3-x/2-\delta}$, with $\delta=0$ (as illustrated in Fig. 1, referred to here as the reference composition). Our prior analysis showed that the reference composition with $\delta=0$ is energetically favorable under a wide range of realistic operating conditions for fuel and electrolysis cells. While SrTiO$_3$ is a typical perovskite wide band gap semiconductor[21-22], Sr$_2$Fe$_2$O$_5$, a MIEC, is a spin-polarized ordered oxygen vacancy compound that undergoes an order/disorder transition at T = 800 C to a disordered, oxygen deficient perovskite phase[23-24]. The cluster expansion provides a computationally efficient mechanism to describe the evolution of properties across this full composition space. The reference composition SrTi$_{1-x}$Fe$_x$O$_{3-x/2}$ is naturally associated with an inherent partial occupancy of the oxygen sublattice in which *x/2* of the usual perovskite oxygen sites are vacant. In our approach, rather than being defects these vacant sites are thought of as natural, intrinsic components of the 'pristine' system.

This methodology enables characterization of the degree of disorder on the Ti/Fe (B-site) sublattice and the O/Vo on the oxygen sublattice as a function of composition and temperature. Consistent with experiment, we do not observe long-ranged ordering, but find that at low temperature short ranged order can be present. Since electronic structure, optical properties, and transport properties are sensitive to the local structure, we perform Monte Carlo simulations to identify realistic configurations for selected compositions. We predict a band gap evolution with Fe content that agrees well with experiment[8]. Together with the electronic structure analysis, we find that the large degree of mixing prevalent in the Fe/Ti sublattice together with the tendency for oxygen vacancies to cluster around Fe atoms helps to spatially distribute the valence band states throughout the STF network and avoid the formation of localized band edge states that would be detrimental for hole transport. Since many MIEC materials of interest may be considered as alloys between perovskites and ordered oxygen vacancy compounds, our study not only gives detailed insights into the structure and electronic properties of alloyed $SrTiO_3/Sr_2Fe_2O_5$ compounds, but also provides a framework for investigation and discovery of other perovskite oxides based solutions for fuel cell applications[1-6].

## 2. Methodology

### 2.1. First-principles calculations

Density functional theory[25-26] (DFT) calculations are used to calculate the total energy and electronic structure of different configurations of the STF alloy. We use the VASP code and the Perdew-Burke-Ernzerhof[27] (PBE) description of the exchange-correlation functional and projector augmented-wave (PAW) pseudopotentials[28]. As DFT-PBE typically does not sufficiently accurately describe localized transition metal states, we include Hubbard U approximations for the Ti 3d and Fe 3d states. Without this, DFT-PBE predicts a band gap of 1.67 eV for $SrTiO_3$ compared to the experimental value of 3.25 eV[2, 4, 10, 21, 29]. Meanwhile, $Sr_2Fe_2O_5$ is metallic according to DFT-PBE while in experiment the band gap is reported as 2.35 eV[8, 30]. To calibrate reasonable U values, we performed benchmark calculations by screening all the U combinations from 0 eV to 8 eV for both Ti and Fe 3d states (see Supporting Information). By comparing the predicted electronic structure with both experimental and hybrid functional predictions, we found that U = 3 eV for Ti and U = 5 eV for Fe provides a reasonable description of both band gaps (2.13 eV for $SrTiO_3$ and 0.95 eV for $Sr_2Fe_2O_5$), density of states, and atomic magnetic moments. The DFT+U approach represents a compromise between accuracy and computational efficiency, necessary when simulating a large number of supercells spanning across different compositions and configurations.

To initialize the configurations, 160 atom supercells of $SrTiO_3$ are first constructed, and then random substitution of the desired concentration of Ti atoms by Fe atoms and corresponding random removal of O atoms is carried out.

**2.2. Cluster expansion modeling and Monte Carlo simulation**

The energetics of different configurations of the disordered $Sr(Ti_{1-x}Fe_x)O_{3-x/2}$ alloy are described using a cluster expansion formulation. The total energy of a particular atomic configuration is expressed as[31-32]:

$$E = \sum_i n_i \varepsilon_i + \sum_{i>j} n_{ij} \varepsilon_{ij} + \sum_{i>j>k} n_{ijk} \varepsilon_{ijk} + \sum_{i>j>k>l} n_{ijkl} \varepsilon_{ijkl} \cdots \cdots \quad (1)$$

Here $\varepsilon_i$, $\varepsilon_{ij}$, $\varepsilon_{ijk}$ and $\varepsilon_{ijkl}$ are coefficients representing energy contributions from each cluster containing monomers made from atom $i$, dimers made from atoms $i$ and $j$, trimers made from atoms $i, j, k,$ and tetramers made from atoms $i, j, k, l,$ respectively.

The integers $n_i$, $n_{ij}$, $n_{ijk}$ and $n_{ijkl}$ refer to the number of times the particular cluster appears in a given sample. In the $Sr(Ti_{1-x}Fe_x)O_{3-x/2}$ configuration space, the disorder appears in two main forms: on the B-site (Ti/Fe) sublattice and on the oxygen sublattice (which can contain an oxygen atom or an oxygen vacancy). As a result, the cluster expansion is built upon the Ti/Fe and O/$V_O$ sublattice only, and Sr atoms are present only as 'spectators' in our model. For all configurations considered, we set a cutoff distance of 8 Å for dimers, trimers and 6 Å for tetramers to describe energetics of the STF alloy (details in ). The energies are determined by fitting across a variety of samples containing different distributions of atoms.

In contrast to allowing full compositional variation as typically carried out in cluster expansions[33-38], the description presented here is restricted to the reference composition $Sr(Ti_{1-x}Fe_x)O_{3-x/2+\delta}$ with fixed $\delta = 0$. This nominally allows both transition metals to maintain, on average, their most favorable oxidation state ($Ti^{+4}$, $Fe^{+3}$) along a smooth transition between $SrTiO_3$ and $Sr_2Fe_2O_5$. Although this is a coarse estimate of the oxygen content as a function of alloy composition (which can be oxidized or reduced relative to $\delta = 0$ depending on the thermodynamic environment), it enables us to develop a cluster expansion across the full composition space spanning between two inequivalent crystal structures of $SrTiO_3$ ($Pm\bar{3}m$) and $Sr_2Fe_2O_5$ ($I2cm$). It additionally allows the fitting of the cluster expansion coefficients to parts of configurational space that are more likely to be sampled in reality, rather than fitting to less realistic stoichiometries.

Once the cluster coefficients are determined, the cluster expansion is then used in Monte Carlo simulations to identify lowest energy configurations and equilibrium configurations of non-zero temperature at a given super cell size [36-37, 39-40]. With the help of the cluster expansion Monte Carlo (CEMC) approach, the electronic structure of $Sr(Ti_{1-x}Fe_x)O_{3-x/2}$ and its temperature dependence accounting for configurational entropy can be directly assessed using representative configurations at different temperatures. We use CEMC to obtain representative configurations at T = 0 K (low energy configuration at a given composition) and higher temperature T = 1000 K. These configurations are used to characterize the alloy's optical and electronic properties using DFT+U. To obtain the representative structure at the high temperature, CEMC simulations are performed to generate the probability distribution of different configurations. From this, the atomic configuration with highest probability is chosen for further analysis (details in Supporting Information).

**2.3 Ordered Structures Used for Comparison**

For the purposes of comparing the electronic structure and optical properties, the solid solution configurations obtained using CEMC are compared to two hypothetical ordered structures. The first is an ordered mixture in which the Fe/Ti and the O/Vo sublattices each exhibit regular order. On the B-site sublattice, Fe atoms at concentration *x* are introduced in an ordered manner; on the oxygen sublattice, oxygen vacancies of concentration *x/2* are introduced in an ordered manner. This gives rise to configurations where oxygen vacancies are adjacent to both Ti and Fe atoms. The second ordered structure is a superlattice composed of distinct local domains of $SrTiO_3$ and $Sr_2Fe_2O_5$. The upper part of the simulation cell is occupied by $Sr_2Fe_2O_5$ and the bottom by $SrTiO_3$ in the appropriate ratio for the given value of $x$. In reality phase separation would occur at much larger length scales, distinct from the superlattices considered here. However, the superlattice reflects the case of minimal Ti/Fe mixing for the given supercell size. Overall, the two hypothetical ordered structures represent extreme cases of (i) regular uniform mixing of Fe atoms in the Ti lattice and oxygen vacancies in the oxygen sublattice, and (ii) the least possible degree of mixing.

## 3. Results and discussion

### 3.1. Cluster expansion training and energy convex hull

The first task is to determine the cluster expansion coefficients and the convex hull for $0 < x < 1$. We generate configurations using a 160 atom supercell that, referenced to the traditional perovskite cubic cell, has lattice vectors oriented along the $[110]$, $[1\bar{1}0]$, and $[001]$ direction. As illustrated in Fig. 1, this supercell can simutaneously capture the structural features of both the cubic $SrTiO_3$ and the distorted $Sr_2Fe_2O_5$ lattice.

We sampled the compositional space of $Sr(Ti_{1-x}Fe_x)O_{3-x/2}$ at seven selected Fe concentrations given by $x = 0.125, 0.25, 0.375, 0.5, 0.625, 0.75, 0.875$. For each, 50 random structures are generated for obtaining the cluster interaction coefficients. We used an iterative approach in which the current cluster expansion coefficients and Monte Carlo sampling are used to predict lowest energy configurations, and these new configurations are then analyzed in DFT and added to the training set to obtain the next set of coefficients. This procedure is repeated until a self-consistent prediction of lowest energy configurations is obtained. For each lowest energy search, we found that $2\times 10^5$ CEMC steps are sufficient for convergence of configurations.

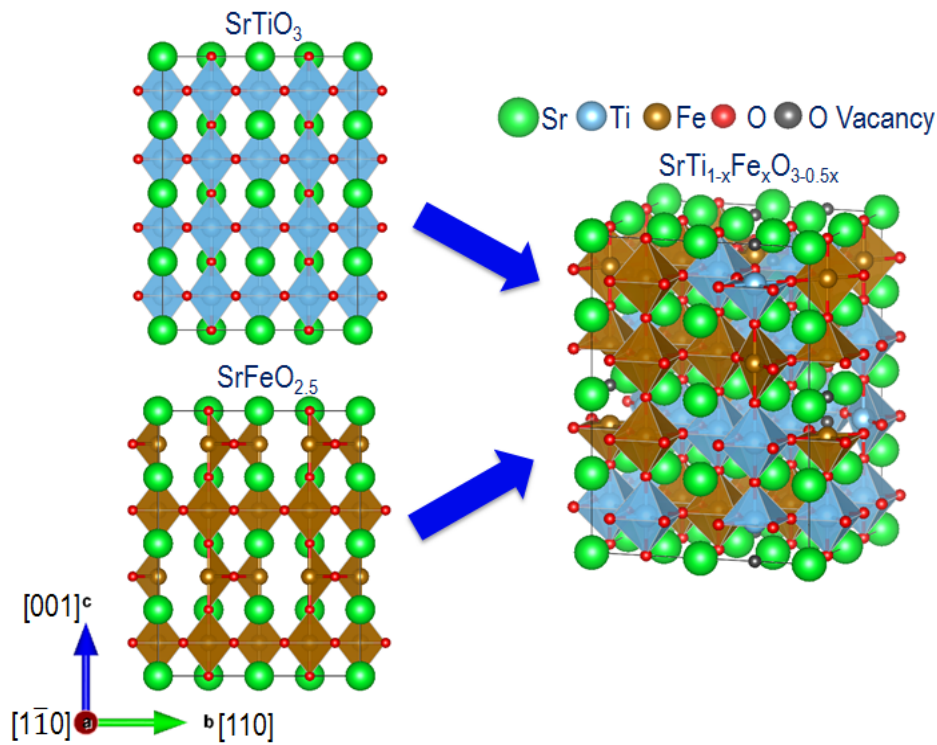

**Fig. 1:** Illustration of the atomic configuration of $SrTiO_3$, $SrFeO_{2.5}$ and $SrTi_{1-x}Fe_xO_{3-0.5x}$ lattices. The $SrTi_{1-x}Fe_xO_{3-0.5x}$ can be regarded as a mix of $SrTiO_3$ and $SrFeO_{2.5}$ with disorder of Fe and Ti cations.

The resulting fit comparing DFT+U to cluster expansion energies is shown in Fig. 2a. A linear least squares fitting of the overall 350 datasets from DFT+U result in an overall root mean squared error (RMSE) of 4.33 meV/atom for the 27 cluster interaction coefficients. We also implemented a recently proposed Bayesian approach[40] to determine the cluster interaction coefficients, which yielded similar results.

To establish the predictive (in addition to descriptive) capability and ensure against overfitting, we carried out a leave-one-out cross-validation analysis. The cross-validation score is expressed as

$$CV^2 = \frac{1}{N_\sigma} \sum_\sigma (H_{mix}^\sigma - \hat{H}_{mix}^\sigma)^2 \qquad (3)$$

where $\hat{H}_{mix}^\sigma$ is the predicted value of mixing enthalpy on the test set $\sigma$ using the cluster interaction coefficients obtained from the training subset. A cross validation score of 4.76 meV/atom is obtained. Therefore, both RMSE and CV scores indicate that our model provides reasonable prediction of energetics with the variation of atomic configurations.

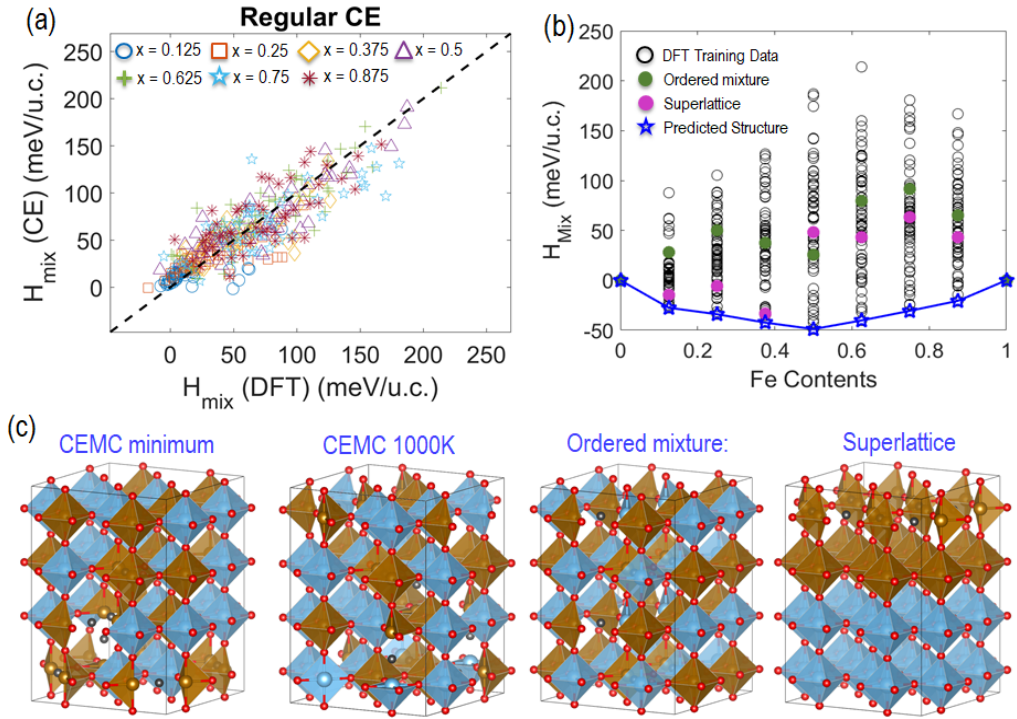

**Fig. 2:** (a) Linear least squares fitting of mixing enthalpy using cluster expansion; 'u.c.' denotes the five atom unit cell of the conventional perovskite lattice. (b) Convex hull showing the lowest energy configurations predicted from Monte Carlo simulation. The training data and two ordered structures are shown for comparison. (c) Atomic configurations of CEMC predicted lowest energy state, CEMC predicted structure at T = 1000 K, and two types of ordered structures. For the convenience of visualization, A-site strontium atoms are not shown.

Figure 2(b) shows the distribution of mixing energies across the full composition range and the convex hull using the lowest energy configurations obtained. The mixing energy is given by:

$$H_{mix} = H_{Sr(Ti_{1-x}Fe_x)O_3} - (1-x) \cdot H_{SrTiO_3} - x \cdot H_{Sr_2Fe_2O_5} \qquad (2)$$

where $H_{Sr(Ti_{1-x}Fe_x)O_3}$, $H_{SrTiO_3}$ and $H_{Sr_2Fe_2O_5}$ are the calculated enthalpy of $Sr(Ti_{1-x}Fe_x)O_{3-x/2}$, $SrTiO_3$ and $Sr_2Fe_2O_5$, respectively. The set of lowest energy configurations across all compositions, denoted by the blue line, yields the curve for the mixing enthalpy. The negative mixing energy demonstrates that $SrTiO_3$ and $Sr_2Fe_2O_5$ mix with each other across the full composition range, consistent with experimental observations[8-9]. Regardless of the alloy composition, a disordered solid solution is energetically more favorable relative to decomposition to $SrTiO_3$ and $Sr_2Fe_2O_5$.

Fig. 2(c) shows representative lowest energy configuration and T = 1000 K configurations obtained using CEMC at the composition of $x = 0.5$. The lowest energy structure contains several Fe-Vo-Fe trimers, which locally resemble the structure of $Sr_2Fe_2O_5$. This is consistent with the suggested presence of short ranged order in STF. At the higher temperature, these clusters are still present but to a lesser extent due to configurational entropy, showing the expected loss of short-ranged order at higher temperature. Also shown in Fig. 2(c) are the two ordered structures that will be used for comparison: first the ordered mixture and the second the superlattice. In Fig. 2(c) the Sr sublattice is not shown since the lattice disorder arises from Ti/Fe cation and O/V$_O$ distribution alone.

From Fig. 2(b), we observe that the energy of the two ordered structures are always higher than the CEMC predicted low energy configuration. This can be understood by comparing the two sets of ordered structures with the CEMC predicted lowest energy structure. Two observations are relevant. First, Ti and Fe cations have a tendency to mix with each other fairly randomly. Second, oxygen vacancies tend to remain adjacent to Fe atoms instead of Ti atoms, and the low energy configurations contain Fe-$V_O$-Fe structures that locally resemble $Sr_2Fe_2O_5$, largely consistent with the maintenance of $Ti^{+4}$ and $Fe^{+3}$ oxidation states. This also explains why forming either an ordered mixture or a superlattice is not energetically favorable. In the ordered mixture, Fe atoms are introduced periodically in the B-site sublattice, and oxygen vacancies are introduced periodically into the oxygen sublattice without regard to the neighboring B-site species. This results in fewer vacancies adjacent to Fe atoms and oxidation states that deviate from $Ti^{+4}$ and $Fe^{+3}$. On the other hand, the superlattice offers the least mixing of Ti/Fe cations. As a result, a structure as suggested by the CEMC in Fig. 2(c) is more energetically favorable because it allows random mixing of Fe/Ti on the B-site sublattice and clustering of oxygen vacancies around Fe atoms.

## 3.2 Electronic structure and Optical Properties of $Sr(Ti_{1-x}Fe_x)O_{3-0.5x}$ alloy

The band gaps of seven selected compositions are calculated using DFT+U and presented in Fig. 3. Results for the two sets of ordered structures are also given for comparison purposes. With the CEMC approach, the electronic structure of $Sr(Ti_{1-x}Fe_x)O_{3-x/2}$ can be directly accessed using the predicted representative configurations. We consider the lowest energy structures found for the seven selected compositions, and use DFT+U to assess the electronic and optical properties. For example, the band gap evolution across the composition space from $SrTiO_3$ to $Sr_2Fe_2O_5$ is presented in Fig. 3(a).

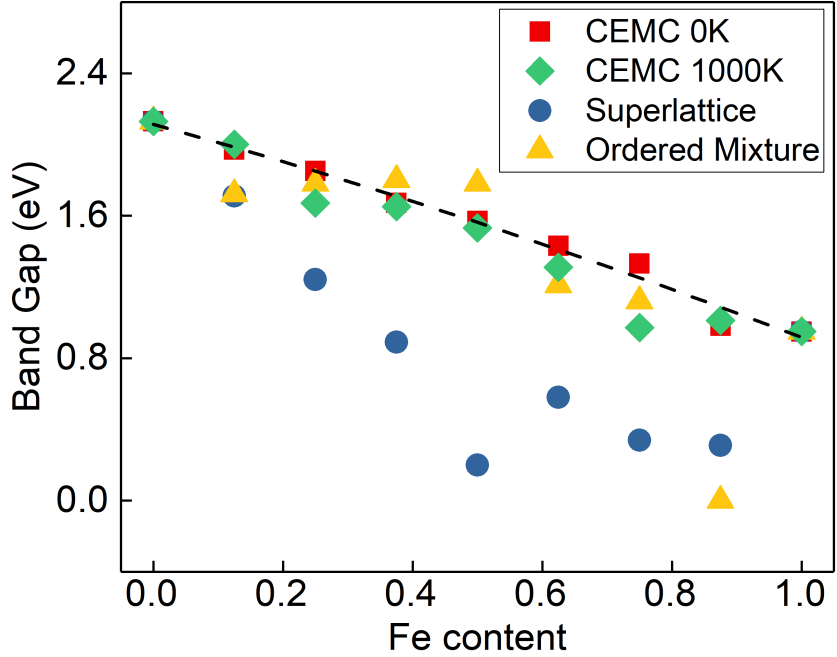

**Fig. 3:** The evolution of band gap with Fe content. For the lowest energy state and T = 1000 K structures, the band gap smoothly decreases with increasing Fe content with little degree of bowing evident. The band gap of the ordered structures are shown for comparison.

Figure 3(a) shows that there are fluctuations in the band gap as the Fe content increases for both sets of ordered structures, without any obvious systematic trends. On the other hand, for the CEMC minimum the band gap appears to decrease smoothly from 2.13 eV to 0.95 eV (DFT+U predicted gaps of $SrTiO_3$ and $Sr_2Fe_2O_5$). Whereas many alloys (Si/Ge for example[41]) exhibit a parabolic dependence of the band gap on the composition according to Vegard's law[8, 42-45], here the bowing parameter (deviation from linear) is quite small, which makes the band gap evolution appear linear. This is consistent with previous experimental results by Rothschild *et al.*[8], suggesting that the cluster expansion approach can be useful for direct estimation of electronic properties in realistic systems. The results for the T = 1000 K configurations are also shown in Fig. 3. Minimal differences are observed compared to the lowest energy structures, with these also showing a smooth decrease in band gap across the composition space.

To elucidate how the atomic configuration affects the electronic structure, we show the projected density of states (PDOS) in Figure 4(a,b) for x = 0.5 and x = 0.875 for the T = 0 K, T = 1000 K, and the two ordered structures. The electronic structure of the lowest energy state and the T = 1000 K configuration are similar to each other for both compositions. Interestingly, however, for x = 0.5 they are both more similar to the ordered mixture while for x = 0.875 they are more similar to the superlattice. For example, at x = 0.5 (Fig. 4a) the PDOS of the lowest state and the T = 1000 K configuration resemble a mixture of the PDOS of $SrTiO_3$ and $SrFeO_{2.5}$ (see supporting information). However the superlattice exhibits a smaller band gap, differences in the atomic contributions to the band edge states, and presence of impurity energy levels. Distinct from the ordered mixture and CEMC structures, the low lying states of the superlattice conduction band exhibit mixed Ti 3d and Fe 3d orbital nature. Meanwhile, occupied mid gap states mainly arising from Fe 3d atomic orbitals are observed above the valence band edge. On the other hand, for x = 0.875 (Fig. 4(b)), the shape and magnetic features of the CEMC structures are more similar to the superlattice, but different from the ordered mixture. In the ordered mixture, the gap closes and the system is metallic, very different from the spin-polarized semiconductor $Sr_2Fe_2O_5$ end-member.

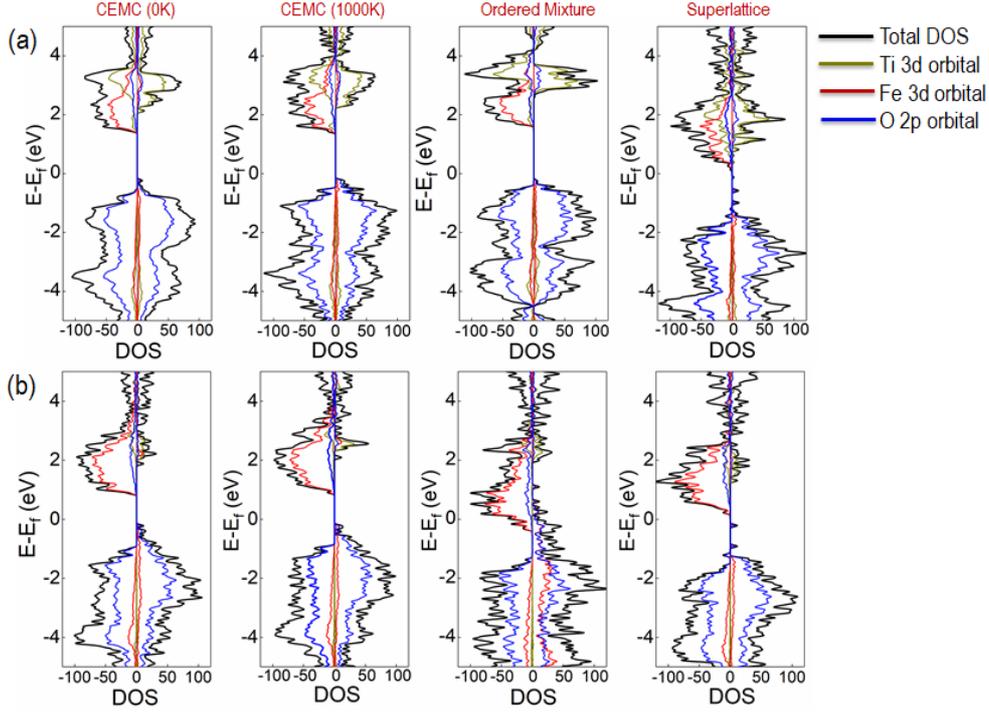

**Fig. 4:** Site and orbital projected density of states (PDOS) of the four configurations of $Sr(Ti_{1-x}Fe_x)O_{3-x/2}$ at (a) $x = 0.5$ and (b) $x = 0.875$.

To understand why for $x = 0.5$ the CEMC structures are more similar to the ordered mixture, whereas for $x = 0.875$ they are more similar to the superlattice, we consider the effects of Ti/Fe ordering as well as oxygen vacancy distribution. At $x = 0.5$, the system has the largest degree of mixing between Ti and Fe atoms. Therefore differences in the degree of Ti/Fe order/disorder dominate the nature of density of states. The superlattice represents the case of minimal Ti and Fe mixing, causing the large observed difference in the PDOS. Additionally the $SrTiO_3$ and $Sr_2Fe_2O_5$ 'interface' in the superlattice plays a significant role in the energy level alignment of Fe and Ti states causing additional discrepancies. On the other hand $x = 0.875$ corresponds to the largest oxygen vacancy concentration considered here. At this composition, the electronic structure is most sensitive to the distribution of oxygen vacancies and correspondingly the oxidation state of each Ti/Fe atom. In the ordered mixture, oxygen vacancies appear adjacent to both Ti and Fe atoms which on average shifts the oxidation states of +4 in Ti atoms more towards +3, and those of the Fe atoms from +3 towards +4. For both B-site species, these shifts lead to realignment of energy levels, and ultimately a different electronic structure.

To understand the influence of Ti/Fe and oxygen vacancy distribution on electronic conductivity, the spatial distribution of the valence band and conduction band edges for the lowest state, T = 1000 K, and two ordered structures for x = 0.5 are illustrated in Fig. 5. The nature of the conduction band minimum (CBM) and valence band minimum (VBM) differ with the variation of ordering in the lattice. Again the lowest energy state and T = 1000 K configuration show similar features. For these, as inferred from Fig. 5(a-d), both the CBM and VBM are spatially delocalized, distributed throughout the supercell, and include contributions from Ti, Fe, and O atoms. As a consequence of Ti/Fe cation disorder, the spatial distribution of band edges is relatively uniform as shown in Fig. 5(i). This is likely to be beneficial for hole or electron transport.

By contrast, for the two ordered structures, both the VBM and CBM are localized to isolated regions of the supercell. For the ordered mixture, we found that the VBM mainly arises from Fe-O networks for Fe atoms coordinated with 5 oxygen atoms (Fig. 5(e)). On the other hand, the Fe with a full octahedral O coordination (Fig. 5(f)) dominates the CBM. As for the superlattice, the VBM is dominated by Fe atoms with octahedral coordination and their first neighboring O atoms (Fig. 5(g)), while the CBM comes from Fe with tetrahedral O coordination (Fig. 5(h)). Such localized states could serve as traps for carriers reducing carrier mobility and transport as suggested schematically in Fig. 5(j). This reveals that the combination of Ti/Fe mixing together with a tendency for vacancy clustering around Fe atoms, when present simultaneously, offers an advantage for electronic transport not exhibited by either of the hypothetical ordered structures.

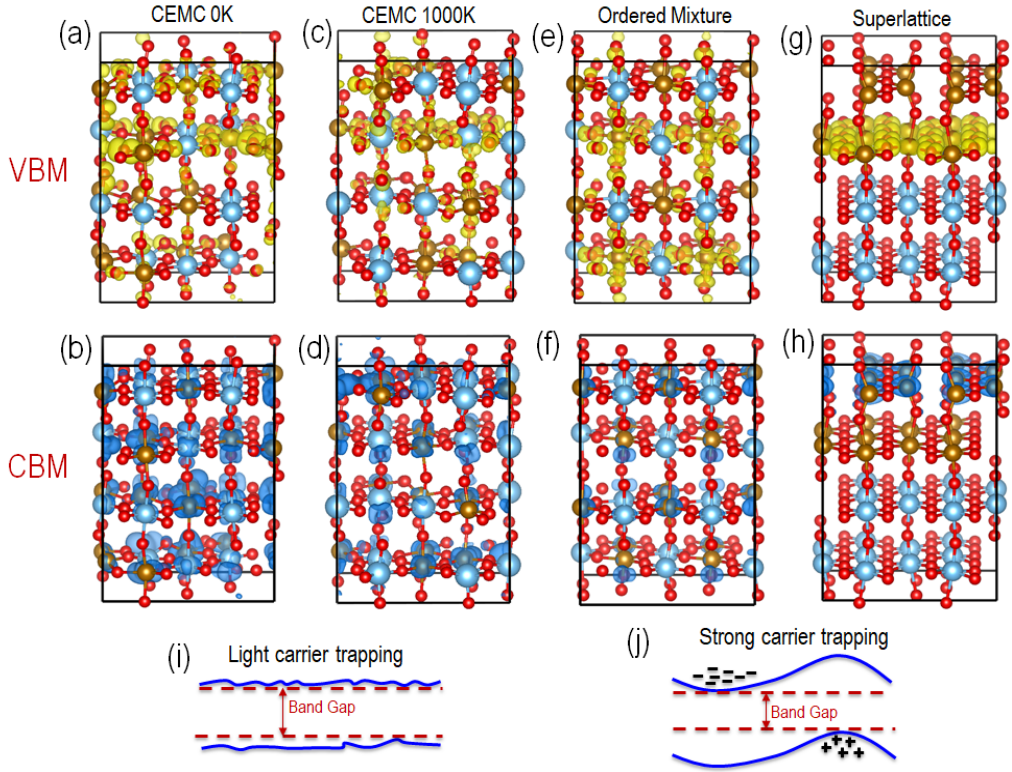

**Fig. 5:** Charge density of the SrTi$_{0.5}$Fe$_{0.5}$O$_{2.75}$ valence band maximums (VBM) and conduction band minimums (CBM). (a) VBM for CEMC predicted minimum energy structure; (b) CBM for CEMC predicted minimum energy structure; (c) VBM for CEMC equilibrium structure at T = 1000 K; (d) CBM for CEMC equilibrium structure at T = 1000 K; (e) VBM for ordered mixture structure; (f) CBM for ordered mixture; (g) VBM for superlattice structure; (h) CBM for superlattice structure. (i) Schematic showing light degrees of carrier trapping due to delocalization of VBM and CBM; (j) Schematic showing strong degrees of carrier trapping due to the spatial distribution of band edges. Strontium atoms are not shown in these figures for visualization purposes.

Lastly, we show the effect of composition on the optical absorption spectrum of the STF alloy. The calculated absorption spectrum based on our DFT+U simulations is given in Fig. 6. Here the CEMC lowest energy state structures are used to obtain the dielectric matrix[46]. The absorption coefficient is formulated from the complex

dielectric function as

$$\alpha(\omega) = (\sqrt{2})\omega\left[\varepsilon^{(1)}(\omega)^2 + \varepsilon^{(2)}(\omega)^2 - \varepsilon^{(1)}(\omega)\right]^{1/2} \qquad (4)$$

where $\varepsilon^{(1)}(\omega)$ and $\varepsilon^{(2)}(\omega)$ refer to the real part and imaginary part of dielectric function respectively. As can be inferred from Fig. 6, the absorption of light becomes gradually enhanced with increasing Fe content. This monotonic variation is consistent with what has been predicted in band gap evolution in Fig. 3.

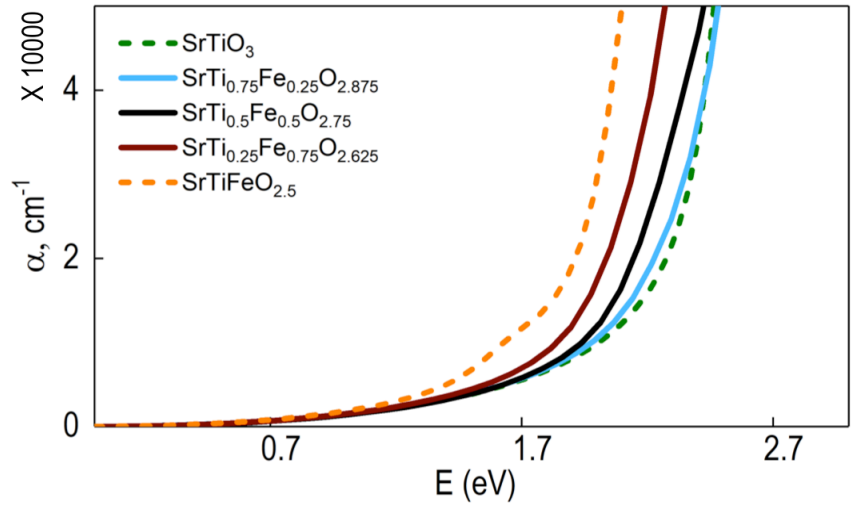

**Fig. 6:** Optical absorption for selected compositions of $Sr(Ti_{1-x}Fe_x)O_{3-x/2}$ alloy for the lowest energy configurations.

## 4. Summary

To conclude, we present a computational framework to consider the effects of composition and order/disorder in the STF MIEC solid solution. Using cluster expansion modeling and Monte Carlo simulations, energetics and realistic configurations of the $Sr(Ti_{1-x}Fe_x)O_{3-x/2}$ can be predicted across the full composition space spanning from $SrTiO_3$ to $Sr_2Fe_2O_5$. We use the framework to generate representative configurations and assess their properties using density functional theory. The analysis reveals the connection between the Ti/Fe cation disorder and oxygen vacancy distribution on the electronic structure. Furthermore, it is found that Ti/Fe cation disorder together with oxygen vacancy clustering around Fe atoms together give rise to spatially delocalized band edge states, which may facilitate electron transport in the lattice. This work not only provides mechanistic understanding of the disorder and electronic structure of $Sr(Ti_{1-x}Fe_x)O_{3-x/2}$, but also suggests a computational strategy for analysis of complex perovskite solutions for fuel and electrolysis cell applications.

## Acknowledgements:

This work is supported by National Science Foundation (USA) under Grant No. 1545907. The computing resources from National Center for Supercomputing Applications in University of Illinois at Urbana–Champaign is also greatly appreciated.

*73* (4), 045112.